\documentclass[pre,aps,twocolumn,showpacs]{revtex4}
\usepackage{epsfig}
\usepackage{bm}
\def\pd#1#2{{\partial #1\over\partial #2}}

\begin{document}

\title{Generalized vorticity in transitional quantum turbulence}

\author{\small  A. Bershadskii}
\affiliation{\small {\it ICAR, P.O.\ Box 31155, Jerusalem 91000, Israel}}

\begin{abstract}
Transition to condensate state in the degenerate gas of bosons is studied using 
the Gross-Pitaevskii equation. 
It is shown that adiabatic invariance of an enstrophy (mean squared vorticity) 
based on a generalized vorticity $curl (|\psi|^2 {\bf v})/|\psi|^2$, controls dynamics of 
the transitional turbulence just before formation of the condensate with its 
tangles of quantized vortices ($|\psi|^2 {\bf v}$ is the weighted 
velocity field defined on the macroscopic wave function $\psi$). 
Scaling of the angle-averaged occupation number spectrum, $N_k \sim k^{-1}$, has been obtained for 
the free developing transitional turbulence for the weak nonlinear (and completely disordered) 
initial conditions. Results of the three-dimensional numerical 
simulations have been used to support the theoretical consideration.   
 
\end{abstract}


\maketitle

{\it Introduction}.  In this paper the term quantum turbulence is related to the stochastic 
dynamics described by the nonlinear Schr\"{o}dinger equation for a macroscopic wave
function that specifies the index of the coherent state (see, for instance, \cite{bs}-\cite{scaling1}). 
Each coherent state evolves (to the leading order) 
along its classical trajectory. The Gross-Pitaevskii equation \cite{gp}, for instance, 
is used for description of such process (in a dimensionless form)
$$
{\rm i} \pd \psi t = - \nabla^2 \psi + |\psi|^2 \psi,
\eqno{(1)}
$$
In particular, this equation gives a description of the formation of Bose-Einstein condensates 
from the strongly degenerate gas of weakly interacting bosons. 

It is believed that the nonlinear Schr\"{o}dinger equation can also describe development of 
the coherent regime \cite{ks},\cite{kss},\cite{ly} at a certain stage of evolution. 
Theoretical considerations (see for instance, \cite{bs}-\cite{st}) show that this regime 
can appear in a long-wavelength region of wavenumber space 
after the breakdown of the regime of weak turbulence: the Kagan-Svistunov (KS-) scenario. 
This is supported by recent tree-dimensional numerical simulations for the weakly interacting Bose gas 
(see for instance \cite{bs}). 
It is expected that there exist three different stages in the time development for the KS-scenario: 
weak turbulence, transitional turbulence and condensate.  A pre-condensate (coherent regime) 
begins its formation in a 
long-wavelength region at the transitional turbulent regime. This transitional turbulent regime 
is the most difficult for theoretical description \cite{bs},\cite{zn},\cite{st}. 
There is a problem to define a continuous vorticity filed in the quantum turbulence due 
to non-rotational nature of the velocity field defined on the wave function given by the Schr\"{o}dinger equation. 
Velocity field defined on the wave functions is a potential field. Vorticity of such field 
vanishes everywhere in a single-connected region. It is believed that in the condensate state itself 
all rotational flow is carried by quantized vortices (the circulation of the velocity around the 
core of such vortex is quantized). This idea turned out to be very fruitful, and recent experiments 
give direct support for existence of such vortices in the condensate \cite{Donnelly},\cite{vn},\cite{sreeni}. 
However, for the transitional turbulence apparent absence of a continuous field associated with vorticity 
is a difficult problem. Moreover, it is clear that the only flux of the particles from the region of lager energies
toward the future condensate in not sufficient for transformation of the completely disordered 
initial weak turbulence into the condensate with its local superfluid order and 
with the tangles of quantized vortices. A certain vorticity-like quantity have to be involved in 
this process just before appearance of the tangles of quantized vortices.

In present paper we study a generalized vorticity defined on the {\it weighted} velocity field. The 
generalized vorticity is not vanishing in the bulk of the flow. Then, we show that adiabatic invariance 
of an enstrophy (mean squared generalized vorticity) 
controls the transitional turbulence just before formation of the condensate with its 
tangles of quantized vortices.\\

First of all, let us recall certain properties of velocity field defined on the wave function. 
In order to calculate the expectation value of velocity one can evaluate the time 
derivative of space displacement $\langle {\bf r} \rangle$:  
$$
\langle {\bf v}\rangle = \frac{d \langle {\bf r} \rangle}{dt}= \frac{d}{dt} \int {\bf r} |\psi|^2 d{\bf x}= 
 \int {\bf r} \pd {|\psi|^2} t d{\bf x}  \eqno{(2)}
$$
Substituting $\pd {|\psi|^2} t$ provided by the Eq. (1) one obtains
$$
\langle {\bf v}\rangle = \int \frac{{\rm i}\hbar}{2m} [\psi \nabla \psi^* - \psi^* \nabla \psi] d{\bf x} \eqno{(3)}
$$
where we have integrated by parts with corresponding vanishing or periodic boundary conditions. The integrand in the right-hand-side of Eq. (3) is the {\it real}-valued distribution of the quantum velocity 
$$
{\bf j} = \frac{{\rm i}\hbar}{2m} [\psi \nabla \psi^* - \psi^* \nabla \psi]  \eqno{(4)}
$$

The real-valued quantum velocity filed itself one can obtain comparing the definition $
\langle {\bf v}\rangle = \int {\bf v} |\psi|^2 d{\bf x}$  
with Eq. (3), that results in
$$
{\bf v} = \frac{{\rm i}\hbar}{2m}~ \frac{[\psi \nabla \psi^* - \psi^* \nabla \psi]]}{|\psi|^2}  \eqno{(5)} 
$$
i.e. ${\bf j}= |\psi|^2 {\bf v}$ is the flux of probability or weighted velocity. 
Representation (5) can be readily transformed into
$$
{\bf v} = -\frac{{\rm i}\hbar}{2m}~\nabla \left[ \ln \frac{\psi}{\psi^*} \right]  \eqno{(6)}
$$
i.e. the quantum velocity (6) is a potential field and $curl~{\bf v} = 0$ everywhere in a single-connected 
region.\\

{\it Generalized vorticity}. Let us define generalized vorticity on the weighted velocity field as 
$$
\omega = \frac{curl~ (|\psi|^2 {\bf v})}{|\psi|^2} = curl~{\bf v}+ \frac{\nabla |\psi|^2 \times {\bf v}}{|\psi|^2}  \eqno{(7)}
$$
First term in the right-hand-side of Eq. (7) is the {\it ordinary} vorticity. 
When there is no the quantized vortices this term is zero due to Eq. (6). In this situation the second term 
in the right-had-side of Eq. (7) determines the generalized vorticity. In this case ${\bf \omega}$ is {\it orthogonal} to the "plane of motion", which is stretched over the vectors: ${\bf v}$ and $ \nabla |\psi|^2$, at any point of the space. In this sense the transitional turbulence can be considered as a locally two dimensional one.

Dynamical equation for the generalized vorticity similarly to the dynamic equation for the density $|\psi|^2$
$$
\pd {|\psi|^2} t = {\rm i} [ \psi^* \Delta \psi - \psi \Delta \psi^*] \eqno{(8)}
$$
does not contain the {\it nonlinear} terms explicitly
$$
\pd {~{\bf \omega} |\psi|^2} t = [\nabla \psi \times \nabla (\Delta \psi^*)] + 
[\nabla \psi^* \times \nabla (\Delta \psi)]  \eqno{(9)}
$$
(all the right-hand-side terms in the Eqs. (8),(9) come from the linear term of the Eq. (1)). 
This cancellation of the 
nonlinear terms in the dynamical equation for the density, (8), results 
(after integration by parts) in the conservation law 
for the total number of the particles: $N = \int |\psi|^2 d{\bf x} =const$. The cancellation of the nonlinear terms in 
the dynamical equation for vorticity, Eq. (9), results in a less strong result. Namely, the enstrophy (mean squared vorticity) turned out to be an adiabatic invariant for the transitional turbulence.\\

{\it Adiabatic invariance and scaling transitional regime}. 
According to the theoretical predictions \cite{ks},\cite{kss} if one starts from a self-similar 
solution of the equation (1) for weak-nonlinear conditions, then the so-called coherent regime \cite{ks},\cite{kss},\cite{ly}, will be developed at a certain stage of evolution. The first stage 
of evolution leads to an explosive increase of occupation
numbers in the long-wavelength region of wavenumber space where the ordering process takes place. 
From the beginning of the coherent stage of the evolution we will call the long-wavelength region 
of wavenumber space as pre-condensate fraction, while the rest of the wavenumber space we will 
call as above-condensate fraction. 

At a certain time, close to the blow-up time of the self-similar solution, 
the coherent regime sets in. After this time the system has a certain transitional turbulent period. 
At the end of this transitional period appearance of a well-defined tangle of 
quantized vortices indicates the final (condensate) stage of the evolution. Therefore, 
generalized vorticity exchange between the pre-condensate and above-condensate fractions in the transitional period 
seems to be crucial for the process of the condensate formation in the KS-scenario. 

In the case of space isotropy it is convenient to deal with an 
angle-averaged occupation numbers spectrum $N_k$ in variable $k = |{\bf k}|$ in the wavenumber space
$$
N =\int N_k dk  \eqno{(10)}
$$
The total number of particles associated with the pre-condensate fraction
is $N^{pc} = \int_{k'< k_c} N_k' dk' $, where $k_c$ is the wavenumber scale separating (approximately)
the pre- and above-condensate fractions (for its determination from numerical simulations see next section). 
Then the total number of particles associated with the above-condensate fraction is: $ N^{ac}=N-N^{pc}$.
From the dynamical conservation law for the total number of particles, $N = const$, we obtain
$$
\frac{dN^{pc}}{dt} =- \frac{dN^{ac}}{dt}  \eqno{(11)}
$$
Then we can define a single exchange rate of the particles number, $\varepsilon_n$ as
$$
\varepsilon_n = \left| \frac{dN^{ac}}{dt} \right|=\left| \frac{dN^{pc}}{dt} \right|  \eqno{(12)}
$$

Let us also consider angle-averaged enstrophy spectrum $\Pi_k$
$$
\langle \omega^2 \rangle = \int \Pi_k dk  \eqno{(13)}
$$
Then, the enstrophy associated with the pre-condensate fraction
is $\Omega^{pc} = \int_{k'< k_c} \Pi_k' dk' $, whereas enstrophy associated with the 
above-condensate fraction is: $ \Omega^{ac}=\langle \omega^2 \rangle -\Omega^{pc}$. 
The intensive particle flux to the long-wavelength region 
of wavenumber space (where the pre-condensate is formed) involves the enstrophy one. 
Unlike the total number of the particles, the average enstrophy $\langle \omega^2 \rangle$ is not 
an exact invariant of the motion. Though, characteristic time scale of the enstrophy exchange between 
the pre- and above- condensate fractions is expected to be much smaller than the characteristic 
time scale of the $\langle \omega^2 \rangle$ evolution  \cite{ks}-\cite{s}. 
Therefore, the $\langle \omega^2 \rangle$ can still 
be considered (approximately) as an 'adiabatic integral' for the exchange process. As it is 
used for the adiabatic processes this statement can be formalized as following:
$$
\frac{\Omega^{ac}}{|d\Omega^{ac}/dt|} \ll \frac{\langle \omega^2 \rangle}{d|\langle \omega^2 \rangle/dt|},
~~~ \frac{\Omega^{pc}}{|d\Omega^{pc}/dt|} \ll \frac{\langle \omega^2 \rangle}{|d\langle \omega^2 \rangle/dt|}
   \eqno{(14)}
$$
(actually, only one of the inequalities (14) is sufficient for further consideration). For the nonlinear 
system the principal problem is {\it invariance} of the characteristic time-scales in (14) to the rescaling 
$\psi \rightarrow \lambda \psi$ (otherwise the inequalities (14) are meaningless). It can be readily shown 
that the cancellation of the nonlinear terms in the 
dynamical equations (8),(9) provides such invariance for the enstrophy based time-scales in Eq. (14). 

Since for the transitional turbulence $\Omega^{ac}$, $\Omega^{pc}$, and $\langle \omega^2 \rangle$ are still of the 
same order, then we obtain from Eq. (14) following inequalities for the rates
$$
\left| \frac{d\langle \omega^2 \rangle}{dt}\right| \ll \left|\frac{d\Omega^{ac}}{dt} \right|,~~~
\left| \frac{d\langle \omega^2 \rangle}{dt}\right| \ll \left|\frac{d\Omega^{pc}}{dt} \right| \eqno{(15)}
$$

From $\Omega^{ac}= \langle \omega^2 \rangle - \Omega^{pc}$ and Eq. (15) we obtain
$$
\frac{d\Omega^{pc}}{dt}\simeq -\frac{d\Omega^{ac}}{dt} \eqno{(16)}
$$
and now we can define a single exchange rate of the enstrophy, $\varepsilon_{\omega}$ in full 
analogy with Eq. (12)
$$
\varepsilon_{{\bf \omega}} =  \left|
\frac{d\Omega^{ac}}{dt} \right|\simeq  \left|
\frac{d\Omega^{pc}}{dt} \right| \eqno{(17)}
$$
i.e. the adiabatic invariance of the enstrophy substitutes its exact invariance in the case of the transitional 
turbulence.

For the weak nonlinearity one can expect that the angle-averaged spectrum $N_k$ 
for sufficiently large $k$ is proportional to $\varepsilon_n$ (cf. \cite{b},\cite{ldn}). 
This can be also valid for the transitional turbulence 
for sufficiently large $k$. Then with the additional dimensional parameter $ \varepsilon_{{\bf \omega}}$ (which has dimension $T^{-2}$) one obtains scaling law for sufficiently large $k$ from the dimensional considerations
$$
N_k \sim \varepsilon_n \varepsilon_{{\bf \omega}}^{-1/2}~ k^{-1}   \eqno{(18)}
$$

External dissipation or forcing can make the estimates (14) invalid and, hence, destroy this scenario (cf 
\cite{df},\cite{no}). 
Though, in the case of a sufficiently weak and linear external dissipation (forcing) the above consideration can 
be still valid. In this case the exact invariance of the total number of particles $N$ can be replaced by its 
adiabatic invariance (similarly to the enstrophy) and the inequalities similar to the Eq.(14) can be made. Due to 
linearity of the dissipation (forcing) these inequalities will be still invariant to the rescaling: 
$\psi \rightarrow \lambda \psi$.   \\

\begin{figure} \vspace{-1.5cm}\centering
\epsfig{width=.45\textwidth,file=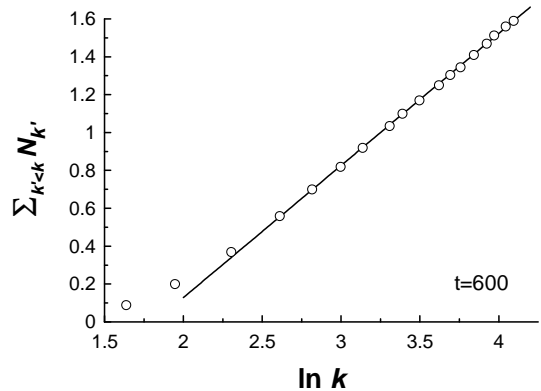} \vspace{-4.5cm}
\caption{The cumulative number of particles $\bar{N}_k  = \Sigma_{k'<k}~ N_{k'}$ against $\ln k$. The data 
are taken from \cite{bs} for $t=600$ (the beginning of the transitional stage). The straight line is drawn in 
order to indicate agreement with Eq. (19). }
\end{figure}

{\it Comparison with numerical simulations}. 
In paper \cite{bs} a large scale {\it three}-dimensional numerical simulations of the Gross-Pitaevskii equation (1) 
was performed in order to reveal all three stages of the evolution from weak turbulence
to superfluid turbulence (the Bose-Einstein condensate) with a tangle of quantized vortices. 
While the observed evolution for $t < 600$ exhibits the well defined self-similar weak turbulence, 
the period $600 < t < 1000 $ was identified in \cite{bs} with the transitional turbulence. In Figs. 1 and 2 
one can see the cumulative number of particles $ \bar{N}_k  = \Sigma_{k'<k}~ N_k'$, 
which shows how many particles have momenta not exceeding $k$ at $t=600$ 
(the beginning of the transitional stage, Fig. 1) and at $t=1000$ (the end of the transitional 
stage, Fig. 2). 
\begin{figure} \vspace{-0.0cm}\centering
\epsfig{width=.45\textwidth,file=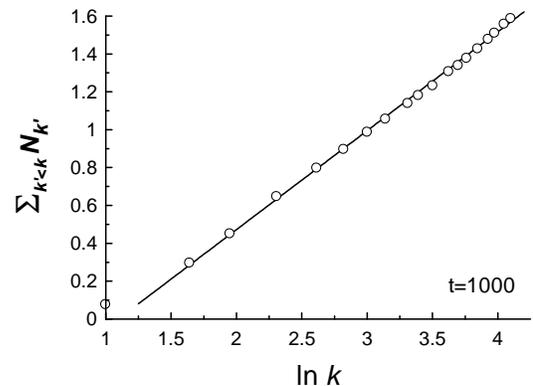} \vspace{-4.5cm}
\caption{As in Fig. 1 but for $t=1000$ (the end of the transitional stage).}
\end{figure}

Using scaling (18) one can estimate this number for sufficiently large $k$ as
$$
\bar{N}_k  = \Sigma_{k'<k}~ N_{k'} \sim  \Sigma_{k_c <k'<k}~ (k')^{-1} \sim \ln (k/k_c)  \eqno{(19)}
$$
where $k_c$ is the wavenumber scale separating (approximately) the pre- and above-condensate fractions.
We use the semi-log scales and the straight lines  in the Figs. 1,2 in order to indicate agreement 
with the Eq. (19). After the formation of the quasi-condensate at 
$t > 1000$ (third stage), the distribution of particles acquires a bimodal shape \cite{bs}. The value $k_c$ 
can be defined at crossing point of the straight line in Figs. 1,2 (indicating Eq. (19)) and 
the horizontal axis.

{\it Conclusions}.  In the above studied KS-scenario the particle flux to the long-wavelength region 
of wavenumber space (where the pre-condensate is formed) initiates and carries the generalized vorticity one. 
At the transition turbulence stage the intensive exchange (transfer) of the particles (generalized vorticity) 
between the pre- and above-condensate fractions of the wavenumber space results in an approximate 
adiabatic invariance of the generalized enstrophy (the average enstrophy dynamics is much slower than 
the exchange dynamics). On the one hand, this adiabatic invariance together with exact conservation law for the total 
number of particles results in a scaling law for the particles number spectrum Eq. (18), on the other hand 
it leads to transition from completely disordered situation at the initial weak turbulence to local superfluid order 
with the tangles of quantized vortices (condensate). The scaling law (18) is a quantitative evidence of this 
scenario. In this scaling law the initial weak nonlinearity is driven by the generalized vorticity exchange 
(under the enstrophy adiabatic invariance) to the condensate state that involves significantly 
different length scales and results in the scaling on the transitional turbulent stage of the condensate 
formation.


\begin{thebibliography}{99}
\bibitem{bs} N. G. Berloff1 and B. V. Svistunov, Phys. Rev. A {\bf 66}, 013603 (2002).
\bibitem{ks} Yu.~Kagan and B.V.~Svistunov, Phys. Rev. Lett. {\bf 79}, 3331
(1997).
\bibitem{svist} B.V. Svistunov, J. Moscow Phys. Soc. {\bf 1}, 373
(1991).
\bibitem{kss} Yu. Kagan, B.V. Svistunov, and G.V. Shlyapnikov, Zh. Eksp. Teor. Fiz. 
{\bf 101}, 528 (1992) [Sov. Phys. JETP {\bf 75}, 387 (1992)].
\bibitem{ks1} Yu. Kagan and B.V. Svistunov, Zh. Eksp. Theor. Fiz. {\bf 105},
353 (1994) [Sov. Phys. JETP {\bf 78}, 187 (1994)].
\bibitem{s} B.V. Svistunov, arXiv:cond-mat/0009368 (2000)
\bibitem{ly} E. Levich and V. Yakhot, J. Phys. A: Math. Gen. {\bf 11}, 2237
(1978).
\bibitem{zn} V.E. Zakharov and S.V. Nazarenko, Physica D, {\bf 201} 203 (2005).
\bibitem{Nazarenko} Y. V. Lvov, S. Nazarenko, and R. West,  Physica D {\bf 184}, 333 (2003).
\bibitem{kopnin} N.B. Kopnin, Theory of nonequilibrium superconductivity
(Clarendon Press, Oxford, 2001).
\bibitem{Davis} M.J.~Davis, S.A.~Morgan, and K.~Burnett, Phys. Rev. Lett.
{\bf 87}, 160402 (2001).
\bibitem{scaling2} M. Kobayashi and M. Tsubota, Phys. Rev. Lett. {\bf 94}, 065302 (2005).
\bibitem{scaling1}T. Araki, M. Tsubota, and S.K. Nemirovskii, Phys. Rev.
Lett. {\bf 89}, 145301 (2002).
\bibitem{gp} V. L. Ginzburg and L. P. Pitaevskii, Sov. Phys. JETP
{\bf 7}, 858 (1958); L. P. Pitaevskii, Sov. Phys. JETP {\bf 13}, 451
(1961); E. P. Gross J. Math. Phys. {\bf 4}, 195 (1963).
\bibitem{st} D. V. Semikoz and I. I. Tkachev, Phys. Rev. D, {\bf 55}, 489 (1997).
\bibitem{Donnelly} C.F.~Barenghi, R.J.~Donnelly,
and W.F.~Vinen (Eds.), {\it Quantized Vortex Dynamics and
Superfluid Turbulence} (Lecture Notes in Physics, Vol. 571),
Springer-Verlag, 2001.
\bibitem{vn} W.F. Vinen, and J.J. Niemela, Quantum turbulence. J. Low Temp. Phys. {\bf 128}, 167 (2002).
\bibitem{sreeni} G. P. Bewley, D. P. Lathrop, and K. R. Sreenivasan, Nature {\bf 441}, 588 (2006).
\bibitem{b} A. Bershadskii, J. Stat. Phys., {\bf 128}, 721 (2007).
\bibitem{ldn} J-P. Laval, B. Dubrulle and S. Nazarenko, Phys. Fluids, {\bf 13}, 1995 (2001).
\bibitem{df} A. Dyachenko, and G. Falkovich, Phys. Rev. E {\bf 54}, 5095 (1996).
\bibitem{no} S. Nazarenko and M. Onorato, Physica D, {\bf 219}, 1 (2006); 
J. Low Temp Phys. {\bf 146}  31 (2007).

\end{thebibliography}
\end{document}